\documentclass[conference]{IEEEtran}
\IEEEoverridecommandlockouts
\usepackage{cite}
\usepackage{amsmath,amssymb,amsfonts}
\usepackage{algorithmic}
\usepackage{graphicx}
\usepackage{textcomp}
\usepackage{xcolor}
\usepackage{subfig}

\usepackage[colorlinks=true,linkcolor=blue,urlcolor=black,bookmarksopen=true]{hyperref}

\usepackage{bookmark}

\def\BibTeX{{\rm B\kern-.05em{\sc i\kern-.025em b}\kern-.08em
    T\kern-.1667em\lower.7ex\hbox{E}\kern-.125emX}}
\begin{document}

\title{Image-Based Feature Representation for Insider Threat Classification \\
}

\author{\IEEEauthorblockN{\textsuperscript{} Gayathri R G}
\IEEEauthorblockA{\textit{School of Information Technology} \\
\textit{Deakin University}\\
Burwood 3125, Australia \\
gradhabaigopina@deakin.edu.au}
\and
\IEEEauthorblockN{\textsuperscript{} Atul Sajjanhar}
\IEEEauthorblockA{\textit{School of Information Technology} \\
\textit{Deakin University}\\
Burwood 3125, Australia \\
atul.sajjanhar@deakin.edu.au}
\and
\IEEEauthorblockN{\textsuperscript{}Yong Xiang}
\IEEEauthorblockA{\textit{School of Information Technology} \\
	\textit{Deakin University}\\
	Burwood 3125, Australia \\
	yong.xiang@deakin.edu.au}

}

\maketitle

\begin{abstract}
Insiders are the trusted entities in the organization, but poses threat to the  with access to sensitive information network and resources. The insider threat detection is a well studied problem in security analytics. Identifying the features from data sources and using them with the right data analytics algorithms makes various kinds of threat analysis possible. The insider threat analysis is mainly done using the frequency based attributes extracted from the raw data available from data sources. In this paper, we propose an image-based feature representation of the daily resource usage pattern of users in the organization. The features extracted from the audit files of the organization are represented as gray scale images. Hence, these images are used to represent the resource access patterns and thereby the behavior of users. Classification models are applied to the representative images to detect anomalous behavior of insiders. The images are classified to malicious and non-malicious. The effectiveness of the proposed representation is evaluated using the CMU CERT data V4.2, and state-of-art image classification models like Mobilenet, VGG and ResNet. The experimental results showed improved accuracy. The comparison with existing works show a performance improvement in terms of high recall and precision values.

\end{abstract}


\section{Introduction}
Drastic increase in the technology left people at high risk of cyber attacks in all domains. E-commerce is increasingly becoming popular on networked devices. The consequences of cybercrimes are frightening. Nowadays, cyber criminals use trending technology to carry out attacks which are difficult to detect. The main goal of cybersecurity is to protect all the resources connected to network ranging from any hardware,  applications like mobile phones and laptops, software and any other data in transit. 

Researchers use data analysis techniques to detect and mitigate security challenges. The use of statistical learning, machine learning, artificial intelligence, natural language processing techniques, etc. has been broadened to solve cybersecurity usecases like detecting malwares and intrusions, phishing, denial of service(DoS) attacks etc. Data analytics is effectively applied in cybersecurity to identify patterns and correlations among events, and to do the behavior analysis to bring out anomalies thereby improving the security system of any organization.

Security analytics is an approach of data processing which combines acquisition of data, aggregation and analysis techniques for security monitoring and threat detection. Security analytics solutions can be applied on large and diverse data sets using machine learning, deep learning and artificial intelligence framework.

Deep Learning(DL), a subset of machine learning, refers to learning the data using a number of layers in a neural network. A shallow network has one hidden layer whereas a deep network has multiple hidden layers that allows it to learn features of the data in a hierarchy. This is because simple features like two pixels are recombined from a layer to the next to make more complex features like a line. In deep learning, input data (features) is passed through more mathematical operations. This is why the training phase in the deep learning algorithms are computationally intensive Hence the DL modeling is achieved widely using GPUs.

Deep learning models are largely employed for security related problems like  intrusion detection, malware analysis. Deep Convolutional Neural Networks(DCNNs or DeepConvNets)  proved to be highly efficient in handling visual data, such as images and videos \cite{image00}. DCNNs take the unprocessed data as input and transforms by processing through a series of basic computational units to get the representations that have useful values for classification in the higher layers.

Insider threat is one of the most popular attacks that happen from the trusted employees or a system or person associated with that organization. These employees may have access to sensitive information and resources. There can be many reasons for the insider attacks like unintentional human error, conflict with co-workers or managers, bribed by competing organizations. Insider threats are mainly targeted at the violation and sabotage of computer systems and the data exfiltration activities. 

As per the recent statistics from the Verizon 2019 Data Breach Investigations report, insiders caused 34\% of all security breaches in 2018 insiders. Cost of Cybercrime study from Accenture \& Ponemon’s 2019 says that average cost of an insider attack rose 15\% from 2018 to 2019. The Ponemon Institute's 2018 Cost of a Data Breach Study claims that an organization requires on average 197 days to identify a security breach and 69 days to contain it.

The insider actions might only leave a small footprint in the digital audit data because attackers know precisely how and where sensitive data resides and is aware of the security solutions implemented in the organization. This is why certain insider incidents are not revealed for a long time. Inexpensive data storage and increased computing power motivates data scientists to acquire and analyze huge volume of audit log data. Effective and efficient data analysis platforms for insider attacks remains an open challenge.

Lot of research happen in the area of insider threat attacks. Ivan Homoliak et.al \cite{survey00} came up with novel categorization on the types of insider threat data available with references to existing works and framework available in the area of insider detection and analysis. Sanzgiri and Dasgupta \cite{dasgupta} classified insider threat detection approaches into nine classes based on the techniques and features utilized in the detection namely : anomaly-based, role-based access control, scenario-based using decoys and honeypots, risk analysis using psychological factors, risk analysis using workflow, improving defense of the network, improving defense by access control and process control to dissuade insiders. Out of these classes, anomaly based detection and user behavior analysis are popularly used. User profiling used in behavior analysis helps in bringing out the abnormal user behavior. It is well established as a primary approach in insider threat detection. The anomaly detection approach focuses on mining anomalous patterns from the normal data or extracted feature vectors.

Different insider threat patterns such as unintentional human errors, misuse of privilege by unsuspecting and malicious users, cyber espionage, etc. aggravate the complexity of insider threat detection. The data used in any learning algorithm used for insider threat attack should be implicated as feature vectors and used for discrepancy or user behavior analysis. In this paper, we propose a method to detect insider threat. In this approach, existing in-depth learning models are adopted for anomaly detection in the context of insider threat detection. The proposed approach is evaluated experimentally and compared to existing approaches, using a publicly available dataset. The results suggest that the proposed approach is effective.

The remainder of the paper is organized as follows. Section 2 provides a general overview on the research context in the insider threat attacks including the various approaches. In Section 3, we explain our proposed approach and the implementation. Section 4 discusses how we used CMU dataset to create images and the pre-trained models to detect insider threats. Finally in Section 5, we conclude the paper by presenting limitations of our work and outlining scope for future work.

\section{Related Work}
Insider threat is a well studied problem that causes drastic damages to the organizations. Researchers work on a plethora of solutions for insider threat detection and prevention.

Zeadally et al \cite{Sherali} present a review of solutions for the insider threat detection with their advantages and shortcomings. As per the paper, the range of existing techniques are categorized into  Intrusion-detection-based approaches, System-call-based approaches, Data-centric approaches, Honeypot approaches, Dynamical-system-theory-based approaches, Anti-indirect-exfiltration approaches and Visualization approaches.

One of the most popular approaches for insider threat problem is framing the problem as an anomaly detection \cite{Berman}. Chandola et al.\cite{Chandola} contributed a a well studied overview of anomaly detection. In this paper, they summarize that anomaly detection methods for sequences with multivariate data is still in the budding stage. Machine learning methods can efficiently tackle the anomaly detection problem and a lot of work happen in that direction. 

Gavai et al.\cite{Gavai} compare supervised classifier approach with an Isolation Forest based unsupervised approach for detecting insider threat using network logs. This work aggregates details of those features that contribute most to the isolation of a data sample in the tree to clearly understand why a user was tagged as anomalous. 

Liu et al. \cite{Liu} proposed a method that could unravel the non-linear relationships in audit data. An ensemble of deep autoencoders is used to detect malicious insider activities by calculating a score from the error resulting from the original data and the reconstructed data. Features are extracted  using a frequency-based approach and these features are used to create datasets for analysis. User behavior can be identified from a set of attributes that are extracted for each category of the audit data log. These features are then given as input to an autoencoder which models the users' behavior. A model is built from each autoencoder for the input features from each audit log. Any deviation of the behavior from the model is treated as an anomaly. Finally, all the models from the individual autoencoder is combined to build a single model that is used to vividly identify the user's behavior pattern. This process takes more time and is not much efficient for a dataset from a different source. One of the limitations to  this approach is that the feature extraction based on frequency do not always give expected outcome. The one-hour interval considered for user behavor study is not enough to identify the usage patterns. 

Noever\cite{David} in his paper tried different families of learning algorithms and concluded that random forest offers a better result with an accuracy of 98\%. The experiments are performed using the CMU CERT dataset and the risk factors are extracted from the data to create the feature vector. They have incorporated the sentiment analysis factors from the email and the website content and file access details.

Meng et al. \cite{Fanzhi} used an approach using Long Short-Term Memory Recurrent Neural Network (LSTM-RNN) and Kernel PCA for the analysis of insiders. The model was built and tested using the CMU Insider Threat dataset v6.2. The feature extraction approach vaguely mentioned the features being used in the method. Performance comparison of the proposed technique was done against popular algorithms such as SVM and Isolation Forest, but it should have been compared with other deep learning models.

In the paper, Lin et al \cite{Lingli} brought forth a hybrid method using Deep Belief Networks(DBN) for feature reconstruction, and One Class SVM (OCSVM) for insider threat detection. In the first step, the features are learned using the DBN model. Then the multi-domain features are re-learned and the hidden features are extracted and trained by the first layer that is visible in DBN followed by each layer of RBM. The last layer is set up with the back propagation network which receives the feature vector generated from RBM and optimizes the parameters of the entire network. Finally, replace the last layer of the network with the trained multilayer structure as a feature extractor. These features are fed to One Class SVM (OCSVM) to train the inside threat detection model which is referred to as the DBN OCSVM model. Each day is divided into intervals of fixed time and calculate the activity frequency from the audit log files. The method achieved an accuracy of 87.79\%.

Yuan et al\cite{Fangfang}, in their paper presented a user behavior anomaly detection based  insider threat detection technique using Deep Neural Network (DNN). User actions sequences are fed to a Long Short-Term Memory (LSTM) which extracts user behavior features and predicts the next user action. A sequence of hidden states of the LSTM model are used to create a feature matrix of fixed-size which is given to the Convolutional Neural Network (CNN) that it as normal or anomaly. The hidden units of the LSTM efficiently captures the temporal behavior patterns. Hence, the long-term temporal dependencies on user action sequence are recorded by LSTM. CERT Insider Threat dataset V4.2. is used for experimentation and the results are promising with an AUC of 0.9449. 

Zhang et al. \cite{ZHANG} focuses on the unsupervised deep learning model DBN. Pre-processing stage includes the collection and analysis of the insider behavior logs to extract the behavior feature in the format of a tuple. It includes the time of the occurrence of the behaviors, the behavior subject, the host that produces the behaviors, and the specific behaviors. All kinds of audit logs have the first three items common in it. The fourth item behavior depends on the behavior types and are more difficult to integrate. 1/N code discretization is applied on the extracted features for data normalization. The deep learning network model DBN uses these features for threat detection. More than one RBM hidden layers with the sigma activation function is used. Back propagation is used to fine tune the network parameters to get an optimized DBN model. CMU CERT dataset is used for method validation.

Chattopadhyay \cite{ScenarioBased} proposed an approach for insider threat detection based on classification of time-series user activities. A features of every single day and its related statistics is computed to construct the time-series features. Since the dataset is highly imbalanced, a cost-sensitive technique for data adjustment is used to randomly undersample the instances belonging to non-malicious class. A deep autoencoder with two layers is used for the classification. The observations show that random forest and deep autoencoder classified the time-series feature vectors with high precision, recall and f-score. Multilayer perceptron gave a higher recall, but it resulted in a low f-score and precision than the other classifiers.

A behavioral analysis framework (BAIT) was proposed by Azaria et al.\cite{Azaria} for insider threat using a semi-supervised classification method that learns from highly imbalanced data. In this work, a one-person game has been designed and 795 players were recruited to play the game on Amazon Mechanical Turk. A few subjects were added to behave maliciously thereby introducing imbalance. Maliciousness is predicted from the way a player plays the game. The paper explains several variations of the BAIT algorithm, amongst which the best approach gave a precision of only 0.07 and a recall of 0.7.

Several surveys\cite{Salem}  \cite{Detecting} has happened in the area of insider threat attack. Salem et al \cite{Salem} mention about different kinds of insider attack as traitors and masqueraders. The paper focused on the challenges of insider threat and categorized most popular techniques into Host-based User Profiling, Network-Based Sensors, Integrated Approaches. One of the main challenges with the research in this direction is the lack of real-world data to study common solutions and general models.

Charlie Soh et al \cite{Charlie} combines deep learning techniques like Gated Recurrent Unit (GRU) and skipgram to construct temporal sentiment details of the employees. The Enron email corpus has been used and anomaly detection is performed on these employee profiles and employees are ranked upon this anomaly score. 

There are a lot of work(\cite{Bhavsar}, \cite{Azaria}) proposed in the field of insider threat detection and prevention using behavioral analysis of people in an organization. 

The class imbalance problem \cite{Zhou} has been considered as a critical issue in machine learning as there are lack of data in various domains and it makes serious impacts on the performance evaluation of learning methods modeled with an assumption of a balanced class distribution. This paper studied the effect of undersampling,oversampling and threshold-moving in the training phase of cost-sensitive neural networks. Soft-ensemble and Hard-ensemble, i.e., the combination of oversampling, undersampling, and threshold-moving using voting mechanisms are also experimented. The main point to be noted is that none of these approaches require any kind of algorithm modification of the neural networks.

The main challenge in insider threat analysis is the need for algorithms and approaches that can efficiently find out the malicious activities with reduced false negatives in optimal time. This is a widely researched problem with a lot work and surveys(\cite{survey1},\cite{survey2},\cite{survey3}) still happening. There are various factors like the data availability, the difficulty in identifying the insider behavior in real-time etc are open issues. Our aim is to come up with a better feature representation of the data that can be applied on any kind of insider dataset.

\section{Proposed Method}
There are existing works that have effectively applied image classification approaches and transfer learning for anomaly detection in the domain of cybersecurity. Transfer learning has been applied to malware classification\cite{malware1} with an accuracy of 98.65\%. In the paper Kancherla and Mukkamala \cite{malware2}, has done malware detection by representing the executable as grayscale images and extracting intensity based and texture based features. This approach achieved an accuracy of 95\% on dataset containing 12000 benign samples and 25000 malware. Tobiyama et al. \cite{malware3} developed a malware detection method that uses RNN to perform feature extraction. The extracted features are represented as an image and a CNN is used for classification to malicious or not and were able to achieve 96\% accuracy. There are papers like \cite{image1},\cite{image2} that proved the image based deep learning analyis for malwares. Recently, Bhodia et. al\cite{transfer1}  used the image based transfer learning for malware analysis. This has motivated us to use image classification approaches for insider threat detection. The behavior of insiders is determined from audit data and mapped to an image. Image classification approaches are applied to these images for anomaly detection. To the best of our knowledge, this is the first work that applies the image classification with the deep learning approach on insider attacks.

\begin{figure}
	\centering
	\includegraphics[scale=0.35]{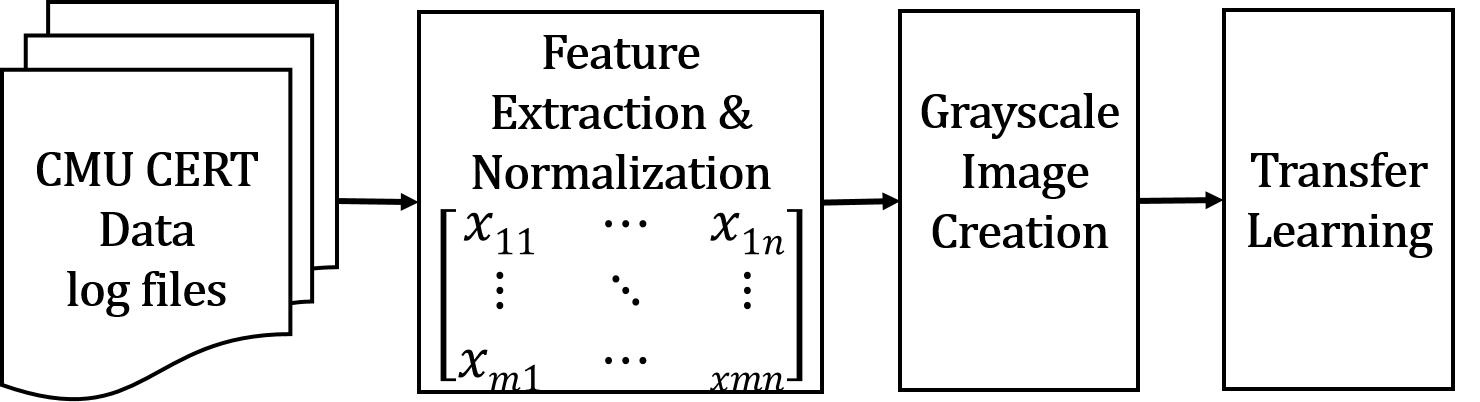}
	\caption{Process Flow}
	\label{fig:flow}	
\end{figure}

The proposed methodology consists of four parts: Pre-processing and Feature extraction, Normalization of the feature vectors, Feature vector representation and Classification. Figure \ref{fig:flow} shows the proposed process flow. 

The feature extraction module reads the log files from the CMU CERT Dataset\cite{CMU}. The log files consists of logon/logoff information, the files handled by the user, the external devices used by the user, email communications sent and received by the user, http details of the browsing history. The logs contain raw information in the form of rows and columns. These logs are used to extract useful features from the data.

\subsection{Feature Extraction}
The \textit{logon.csv} is used to extract the information of the computer usage, the session information, the number of sessions within and outside office hours etc to understand the computer usage pattern on each day. This information helps to understand the users access patterns on the computer in the organization. The \textit{file.csv} gives details about the files copied or read or written by the user on each date. This information can be used to analyze the file access patterns of the user. \textit{Device.csv} log file contains the external device access details of the user of the organization for each day. It helps in finding the users access to any device within and the outside office hours. Usage of the devices outside office hours is suspicious activity. The \textit{email.csv} file contains information about the emails sent and received by the employee in and outside the office hours and the recipient list. The recipient list helps in identifying any external entity who received mail from him. The \textit{http.csv} is about the browsing information of the user in the organization. 
\begin{figure}
	\centering
	\includegraphics[scale=0.27]{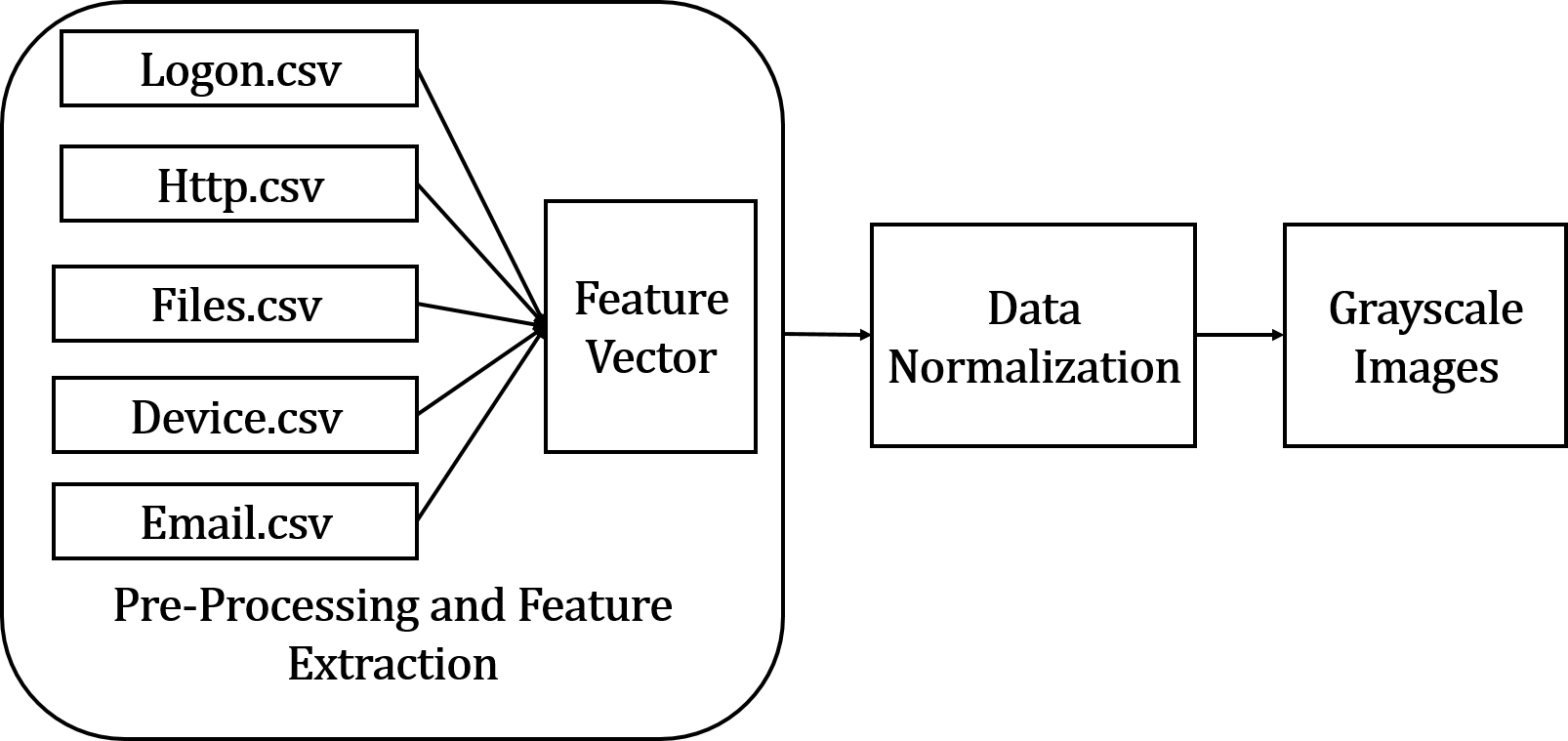}
	\caption{Images creation from Audit Files}
	\label{fig:ImageRepresentation}	
\end{figure}

Features are extracted from all the log files and a feature vector is constructed. All the events in the log files are used to construct the feature vector. The feature extraction mentioned in the paper \cite{ScenarioBased} are used in this work. In the paper \cite{ScenarioBased}, 20 features are mentioned. They use a scenario based approach using the ground-truth information available for malicious activity. In our work, we use the CMU data and extract 20 features for feature representation. L1 to L9 features are extracted from the logon.csv file and it gives the daily computer usage pattern. E1 to E5 gives the email communication details of the user. F1 gives the file access of the user. D1 to D3 are the attributes that explain about the external device usage of the employees. H1 to H3 is used to identify the web access patterns of the user. All these features together contribute to the feature set of the classification task. 

The features contain numeric values. The data distribution into various bins might give varying results in the model evaluation. Min-Max normalization is performed to bin the data in a range 0 to 1 for each date. This date wise normalization is done to ensure that the features are in the same range per day.

The event logs do not contain the malicious user details. The CERT dataset provides the ground truth to mark the users as malicious or non-malicious. The ground truth contain the details of users and the malicious events that has happened with the date and time. The class distribution is highly imbalanced. Majority of the classes belong to non-malicious as the insider attacks happen not so frequently. So the imbalanced data needs to be modified to get better results. Here, we use the random undersampling mechanism since the number of non-malicious instances are very high than the malicious ones. The undersampling is performed with various ratios and the evaluation is performed. 

\subsection{Image-based Feature Vector Representation}
\begin{figure}
	\centering
	\includegraphics[scale=0.3]{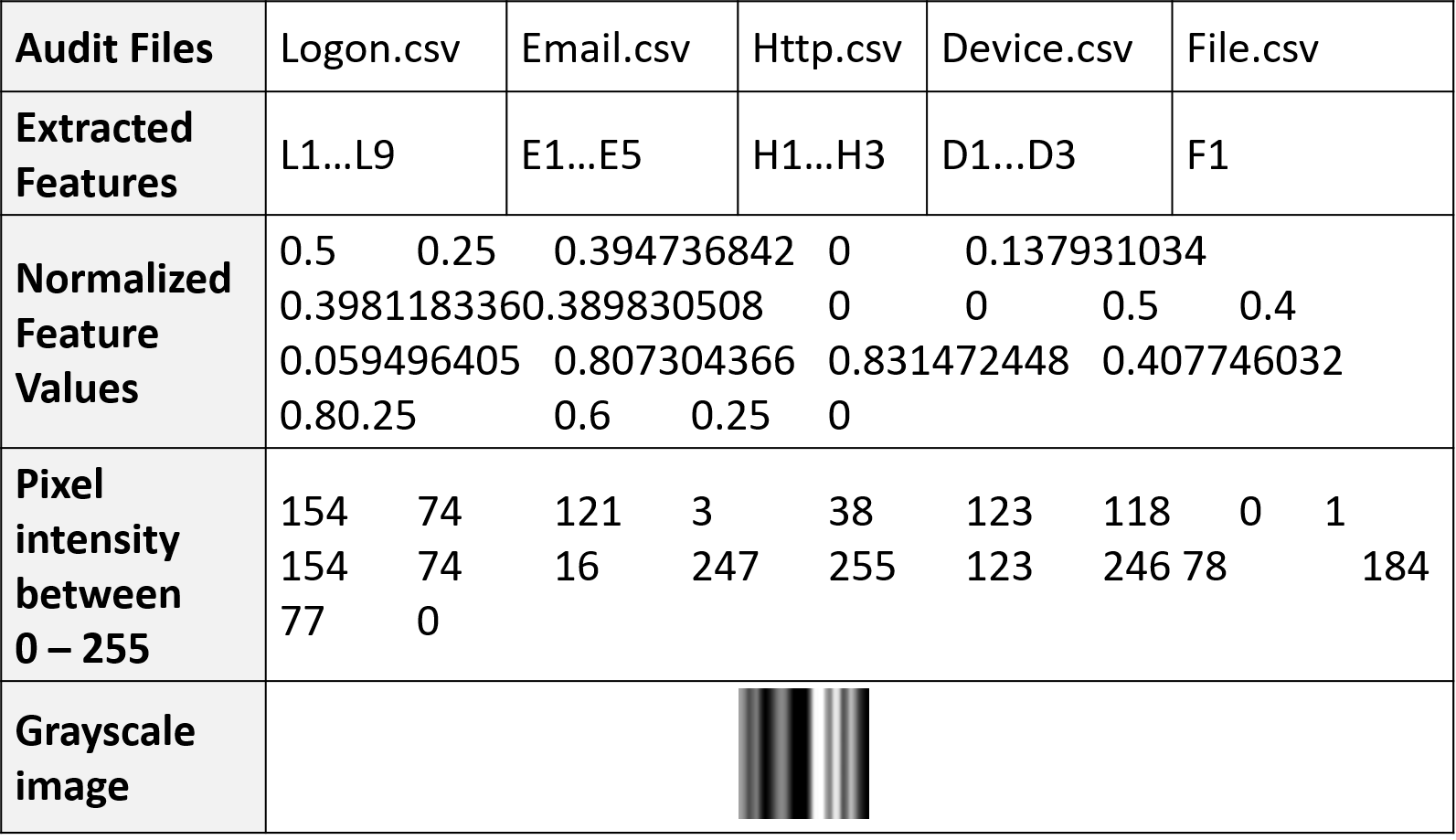}
	\caption{Image creation sequence}
	\label{fig:createimage}	
\end{figure}

The features extracted from the log files are represented as grayscale images and the classification is done to get the malicious and non-malicious users. Figure \ref{fig:createimage} depicts the image creation process from CMU Cert data. The audit files from the CMU dataset are pre-processed to obtain in the feature vector of each user for each day.Each vector is of size 1 x 20. This helps in getting the system usage pattern of each user on a daily basis. It is represented as a grayscale image where each pixel is an attribute that ranges from 0-255. The images are reshaped to a size of 32x32 for training. Each user's daily log pattern is represented as the grayscale images.

\subsection{Transfer Learning for Anomaly Detection}
Transfer learning \cite{transfer} is a research direction in machine learning that focuses on retaining knowledge obtained during  one problem solving and re-applying it to another similar problem. Transfer learning is commonly used with predictive modeling problems that use image data as input. One or more layers from the trained model are then used in a new model trained on the problem of interest.

A pre-trained network performs transfer learning by adding a few dense layers towards the end of the pre-trained network. The learning process aims at finding the most suitable combination of these already learnt features and see how it helps in identifying the samples in our dataset. This makes the training phase fast and requires less data than training a DCNN from the scratch.

Transfer learning is mainly used to eliminate the overhead in training. Transfer learning models can be classified into four: instance-based, parameter, feature-representation and relational-knowledge transfer learning \cite{transfer}. Out of these classifications, feature-representation-transfer learning is used for obtaining a better feature representation for the target domain. In the proposed work, feature-representation 	transfer learning is most suitable. A CNN pre-trained on the ImageNet is used, where the top layers are replaced with the classification head. There is remarkable reduction in the training time, hence pre-trained models are preferred wherever applicable.

\subsection*{Network Architecture}
Deep learning Convolutional Neural Network models are popularly used in image classification and recognition. In order to train and test the images, each input image is passed through a series of convolution layers with filters, pooling, and fully connected layers. We have used the most popular pre-trained models like VGG16, Inception and Mobilenet.

Feature extraction is being done in the base model. As we have a different dataset which is not much similar to the data that is used for the pre-training of the original model, we do the customization of the network and re-train the model. The base model uses the different layers of the pre-trained network to do the feature extraction. In this stage the pixels of the input image are converted into features. Then these features are passed onto classification layer. The global pooling layer does a more intense reduction in dimensionality, such that a tensor of dimension \textit{h×w×d} is reduced to a dimensions \textit{1×1×d}. Once the features are extracted, it is passed to the fully connected layer and the last layer that does the prediction in terms of the probability.

\subsection*{Input images}
We created grayscale images from the log files. The pre-trained network is used for transfer learning. The image size really matters in transfer learning  as the image input size is 224 x 224 etc. Since our feature vector is of small size, we update the input shape dimensions to accept images with the dimensions  of our data and has used an image size of 32x32. If the input dimensions are too large, your network might fail to achieve reasonable accuracy as there are no enough layers in the network to learn the features. If the dimensions of the input image are too small, then the neural network naturally reduces volume dimensions during the forward propagation and then effectively run out of data.

\begin{figure}
	\centering
	\includegraphics[scale=0.6]{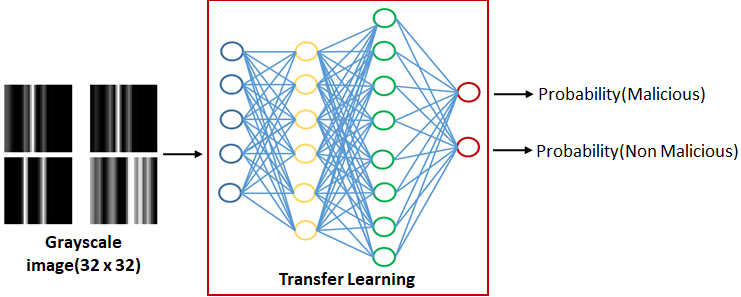}
	\caption{Transfer Learning}
	\label{fig:train}	
\end{figure}

Figure \ref{fig:train} gives an overview of the training phase in the transfer learning. The grayscale images are passed to the DCNN to issue predictions. The network does the prediction using the probability. The error in the prediction is calculated using the loss function.

\section{Experimental Results}
Effectiveness of the proposed method is implemented in Keras with Tensorflow as backend, an open source software platform from Google. In the following section, we detail about the data used parameters tuned,and present the results and analysis.  

\subsection{Dataset}
We use CMU CERT insider threat dataset v4.2\cite{CMU} to simulate the proposed method. CMU CERT data is a benchmark data for insider threat detection with a set of event logs generated from the computer network of a simulated organization. We used the five event files: logon/logoff activity, http data, email communications, file operations, and usage of external storage device. 1000 users generate 3320452 events (log lines) during a period of 501 days. Table \ref{data} gives the overview of the data.

\begin{table}[htbp]
	\caption{Dataset Details}
	\begin{center}
		
		\begin{tabular}{||c c||} 
			\hline\hline
			Total instances & 330452\\ [0.5ex] 
			\hline
			Malicious & 1364\\
			\hline
			Non-Malicious  & 329088\\[1ex] 
			\hline
			
		\end{tabular}
		\label{data}
	\end{center}
\end{table}
\subsection{Random Sampling of Training Data}
In the CMU dataset, the number of malicious instances are very less when compared to that of the non-malicious ones. This shows that the data is highly imbalanced in terms of the class distribution. Imbalanced data typically refers to the data used by the classification tasks where the classes are not distributed equally. For instance, in a binary classification with 100 samples in which 80 samples are labeled as Class 1, and rest of the 20 samples are labeled as Class-2. In this example of an imbalanced dataset, the ratio of Class 1 to Class 2 samples is 4:1.

Imbalanced datasets can be dealt with various strategies like adapting the classification algorithms or balancing classes in the training data before providing the data as input to the learning algorithm. The sampling technique is applied to either add more samples to the minority class referred to as oversampling or remove the samples from the majority class referred to as undersampling. The main idea is to achieve a better balance in the data sample for all the classes.

Undersampling means randomly selecting a subset of samples from the instances with majority class to avoid its influence from dominating the  algorithm learning process. The most common method for doing this is resampling without replacement. In the proposed method, we have used undersampling to make the data balanced to some extent. The original data has an imbalance ratio of 241 : 1. We have used sampling ratio of 150:1 to undersample the majority class instances. On undersampling, only the majority class instance count changes. The minority class instances remain the same. The original dataset is split into 75\% training and remaining is used as test set.

\subsection{Pre-trained CNN Model}

The pre-trained models use two methods to build new models: Feature extraction and Fine-tuning. Having learned a better representation of features from the large Imagenet dataset, the models can act as a feature extractoe for our dataset as well. Though our images are different from the imagenet data, the model should still be able to extract the relevant features from it based on the principles of transfer learning. Load the pre-trained weight trained in the Imagenet data. No weight update is performed in the feature extraction, hence freeze all the lower convolutional layers. Add top layers needed for classifying the classes in the dataste being trained. Here, we train the last layers by adding a global pooling layer to reduce the image dimension, a fully connected layer to convert the features into a single prediction per image and the last  classification layer and train the top layers. The final output is the probabilities for each of the classes in our dataset. The accuracy obtained from the feature extractor model is not as expected. 

\begin{figure}
	\centering
	\includegraphics[scale=0.4]{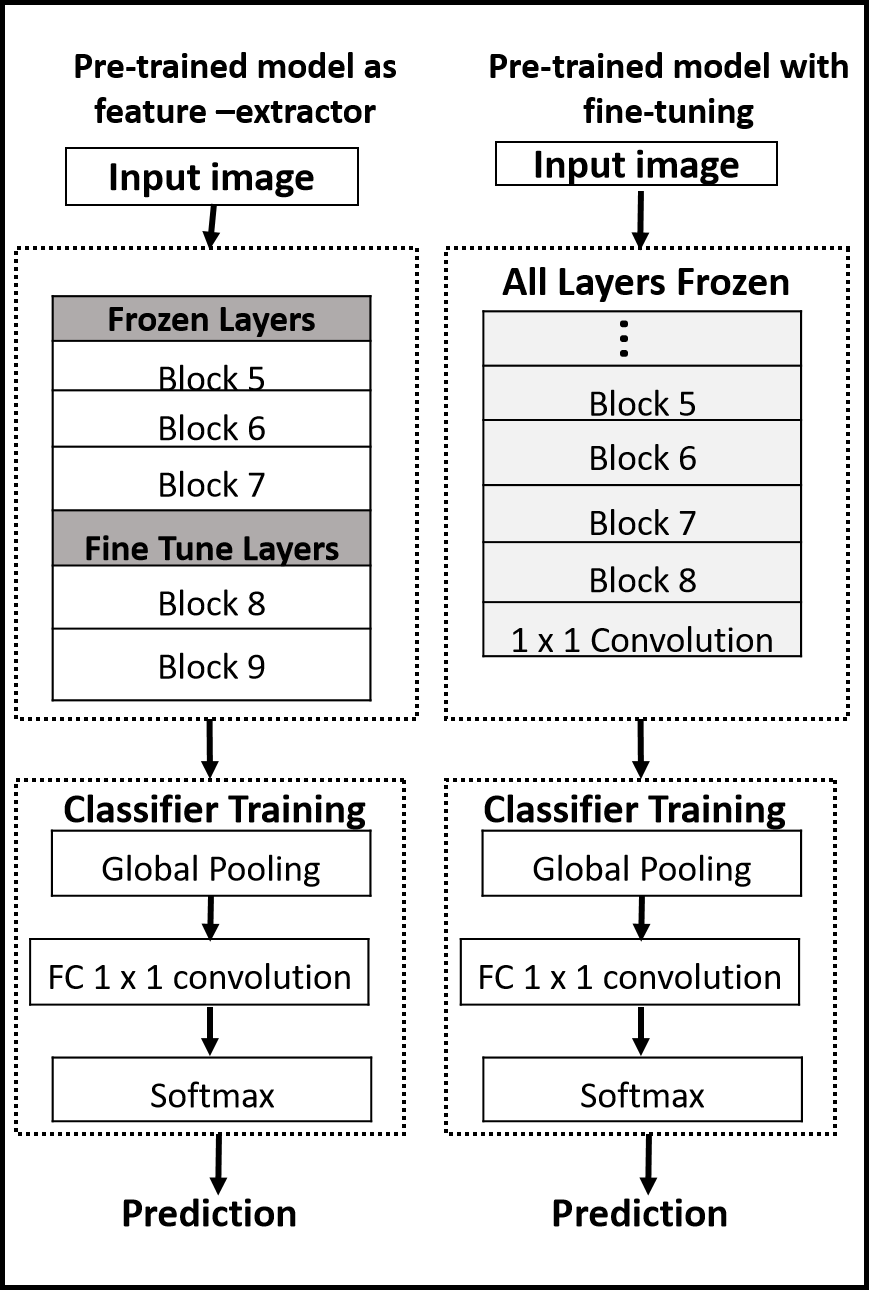}
	\caption{Transfer Learning for new Classes}
	\label{fig:custom}	
\end{figure}

In the feature extraction, we trained a few top layers of the base model. During the training, the weights of the pre-trained network were not updated. Fine-tuning can be applied to increase the performance. Here, a few other layers will be trained by updating the weights while training. The training process will tune the weights to match the features in our dataset. Usually the higher layers of the convolutional networks are more specialized. The lower layers learn generic and simple features that can be generalized to most of the images. Moving up the network, the features become more specific to the dataset on which the model is being trained. The main goal of fine-tuning is to use the specialized features with the desired dataset, rather than overwriting the generic learning.

Figure \ref{fig:custom} gives the details of the network architecture being used for feature extraction and fine tuning. The proposed approach creates a base model that is pre-trained on the ImageNet dataset, a large dataset of 1.4M images and 1000 classes of web images. In our dataset, we have only two classes. In this case, the last layer of the pre-trained model is not helpful. In order to customize the model for the binary classification, load a network without the classification layers at the top. The number of neurons in the last layer of the network should be same as the number of classes to be identified in the dataset. So the 1000 neuron layer is discarded and added our own last layer for the network. In our last layer, we have two neurons as there are two classes(Malicious and Non malicious) with the softmax activation.

The weight is initialized with the imagenet weight used in the training of the original model. Among various Stochastic Gradient descent(SGD), we used the Adams optimizer. Since there are only two class variables, we used the binary crossentropy logarithmic loss function. In a classification task, Softmax function is applied to classify an image with probabilistic values ranging between 0 and 1.

\subsection{Training and Testing}
Table \ref{tab1} shows the number of instances in the feature matrix after normalization. Out of 330452 samples, 1364 are malicious. The undersampling is applied on the 75\% training data, i.e on the 247838 instances.Upon undersampling the majority class instances, the number of instances in majority malicious class reduced to 153900.
\begin{table}[htbp]
	\caption{Training and Test Data}
	\begin{center}
		\begin{tabular}{|c|c|c|c|}
			\hline
			\textbf{CMU}&\multicolumn{3}{|c|}{\textbf{Class Wise Data}} \\
			\cline{2-4} 
			\textbf{Data} & \textbf{\textit{Malicious}}& \textbf{\textit{Non-Malicious}}& \textbf{\textit{Total}} \\
			\hline
			Total& 1364 & 329088 & 330452 \\
			\hline
			Training(Sampled)& 1026 & 153900& 154926 \\
			\hline
			Testing& 338& 82276& 82614\\
			\hline
		\end{tabular}
		\label{tab1}
	\end{center}
\end{table}

\subsection{Results}
The performance evaluation of the model has been done using the CMU data transformed to grayscale images. The weights used in the original pre-trained models are used, but with varying hyper parameters. As it is a binary classification problem, the accuracy, precision and recall are used to compare the results. Table \ref{performance} gives the comparison of the proposed approach with the other existing works.

\begin{table}[htbp]
	\caption{Experiment Results}
	\begin{center}
		\begin{tabular}{||c c c||} 
			\hline\hline
			Method & Precision & Recall\\ [0.5ex] 
			\hline
			BAIT \cite{Richardson} & 51.44 & 82.09\\
			\hline
			Modified Isolation Forest\cite{Gavai2} & 51.44 & 82.09\\
			\hline
			Deep Auto Encoder\cite{ScenarioBased} & 50.42 & 90.25\\ 
			\hline
			LSTM-RNN\cite{Fanzhi} & 95.12 & NA\\ 
			\hline
			Proposed Method3 & 99.32 & 99.32\\[1ex] 
			\hline
		\end{tabular}
		\label{performance}
	\end{center}
\end{table}

\subsection{Discussion}
We proposed the image based insider threat analysis. In this work, we used the CMU CERT insider data to demonstrate our results. The preliminary results obtained clearly shows the improvement in the precision and recall of the existing approaches. The existing approaches used different feature extraction and representation, whereas we have used the feature representation as images. The CNN has trained the data to understand the usage patterns of the user and tries to find the anomalous usage. 

The classification process classified the users as malicious and non-malicious. Our approach achieved an accuracy of 99\% when compared to other techniques. The feature extraction process is not simply frequency based, it helped in identifying the access patterns of various resources like computer, files, external device usage etc of the users on a daily basis. Still there are chances of improvement to the pre-processing and representation used here.

\section{Conclusion and Future Work}
Recent research had shown that deep learning can learn effectively and efficiently from very complex mappings between inputs and outputs directly from the data. It was proven successfully in numerous ﬁelds like speech recognition, image recognition, Natural Language Processing (NLP), etc. But, the huge number of hidden neurons and layers used in Deep Neural Networks end up in computationally intensive vector and matrix operations involving millions of parameters, that needs high-performance computing resources. 
Moreover, it is impractical to collect labeled data samples in many real-world domains to train a network from the scratch. In such cases, a pre-trained deep learning model can be used with ﬁne-tuning it to adapt to the required task with available data and it has resulted in successful results.
In this paper, we studied and evaluated the deep transfer learning technique, to use deep learning models trained on ImageNet and transfer their learning capability to perform insider threat detection from the grayscale images created out of the CMU CERT audit data logs.
The following are some remarkable findings:
\begin{itemize}
	\item Image based anomaly detection can be applied on security related issues. 
	\item The grayscale images used on the pre-trained DNN models gave high precision and recall when compared to the existing works.
\end{itemize}
There is a vast scope for future work on image-based security analysis. Various other attacks can be framed as image or pattern recognition problems and apply the DNN to solve the same. Additionally, more experiments are needed helpful to better portray this effect and prove the success of image-based analytics.

\end{document}